\PassOptionsToPackage{x11names}{xcolor}

\documentclass[sigconf]{acmart}

\copyrightyear{2025}
\acmYear{2025}
\setcopyright{cc}
\setcctype{by}
\acmConference[CCS '25]{Proceedings of the 2025 ACM SIGSAC Conference on Computer and Communications Security}{October 13--17, 2025}{Taipei, Taiwan}
\acmBooktitle{Proceedings of the 2025 ACM SIGSAC Conference on Computer and Communications Security (CCS '25), October 13--17, 2025, Taipei, Taiwan}\acmDOI{10.1145/3719027.3765115}
\acmISBN{979-8-4007-1525-9/2025/10}

\usepackage{wrapfig}
\usepackage{url}
\usepackage{graphicx}
\usepackage{framed}
\usepackage{enumitem}
\usepackage{quoting}
\usepackage{listings}
\usepackage{alloy}
\usepackage{amsthm}
\usepackage{tcolorbox}
\usepackage{hyperref}
\usepackage[hyphenbreaks]{breakurl}
\usepackage{xurl} 
\usepackage{booktabs}
\usepackage{multirow}
\usepackage{subcaption}

\renewcommand\UrlFont\ttfamily\small

\hypersetup{
    colorlinks=true,
    linkcolor=blue,
    filecolor=magenta,      
    urlcolor=blue,
    pdftitle={Overleaf Example},
    breaklinks=true,
    pdfpagemode=FullScreen,
}

\newcounter{fqpcounter}
\newcommand{\fqp}[1]{
  \stepcounter{fqpcounter}
  \par\smallskip\noindent$\triangleright$~FQP~\thefqpcounter.~\textit{#1:}
}

\newcounter{srpcounter}
\newcommand{\srp}[1]{
  \stepcounter{srpcounter}
  \par\smallskip\noindent$\triangleright$~SRP~\thesrpcounter.~\textit{#1:}
}

\newcounter{bpcounter}
\newcommand{\bp}[1]{
  \stepcounter{bpcounter}
  \par\smallskip\noindent$\triangleright$~BP~\thebpcounter.~\textit{#1:}
}

\newcounter{upcounter}
\newcommand{\up}[1]{
  \stepcounter{upcounter}
  \par\smallskip\noindent$\triangleright$~UP~\theupcounter.~\textit{#1:}
}

\newcommand{\highlightbox}[2]{
\begin{tcolorbox}[fonttitle=\bfseries\small, colback=gray!20, colframe=gray!90, title=#1]
\small
{#2}
\end{tcolorbox}
}

\definecolor{codegreen}{rgb}{0,0.6,0}
\definecolor{codegray}{rgb}{0.5,0.5,0.5}
\definecolor{codepurple}{rgb}{0.58,0,0.82}
\definecolor{backcolour}{rgb}{0.95,0.95,0.92}

\lstdefinestyle{mystyle}{
    backgroundcolor=\color{backcolour},   
    commentstyle=\color{codegreen},
    keywordstyle=\color{blue},
    numberstyle=\tiny\color{codegray},
    stringstyle=\color{codepurple},
    basicstyle=\ttfamily\scriptsize,
    breakatwhitespace=false,         
    breaklines=true,                 
    captionpos=b,                    
    keepspaces=true,                 
    numbers=left,                    
    stepnumber=1,    
    numbersep=5pt,                  
    showspaces=false,                
    showstringspaces=false,
    showtabs=false,                  
    tabsize=4,
    frame=lines,
    xleftmargin=5pt,    
    columns=fixed,
}

\newcommand{\point}[1]{\par\smallskip\noindent\textbf{#1.}}

\colorlet{shadecolor}{LavenderBlush2}
\usepackage{lipsum}
\newenvironment{shadedquotation}
 {\begin{shaded*}
  \quoting[leftmargin=0pt, vskip=0pt]
 }
 {\endquoting
 \end{shaded*}
}

\begin{document}

\title{Towards a Formal Foundation for\\ Blockchain ZK Rollups}

\author{Stefanos Chaliasos}
\orcid{0000-0001-5414-4120}
\affiliation{%
  \institution{Imperial College London}
  \city{London}
  \country{UK}}
\affiliation{%
  \institution{zkSecurity}
  \city{New York}
  \country{USA}}
\author{Denis Firsov}
\orcid{0000-0003-1267-7898}
\affiliation{%
  \institution{Input Output}
  \city{Tallinn}
  \country{Estonia}}
\affiliation{%
  \institution{Tallinn University of Technology}
  \city{Tallinn}
  \country{Estonia}}
\author{Benjamin Livshits}
\orcid{0000-0002-4921-8452}
\affiliation{%
  \institution{Imperial College London}
  \city{London}
  \country{UK}}

\renewcommand{\shortauthors}{Stefanos Chaliasos, Denis Firsov, and Benjamin Livshits}
\renewcommand{\shorttitle}{Towards a Formal Foundation for Blockchain ZK Rollups}

\begin{abstract}
Blockchains like Bitcoin and Ethereum have revolutionized digital transactions, yet scalability issues persist. Layer 2 solutions, such as validity proof Rollups (ZK-Rollups), aim to address these challenges by processing transactions off-chain and validating them on the main chain. However, concerns remain about security and censorship resistance, particularly regarding centralized control in Layer 2 and inadequate mechanisms for enforcing these properties through Layer 1 smart contracts. In their current form, L2s are susceptible to multisig attacks that can lead to total user funds loss. This work presents a formal analysis using the Alloy specification language to examine and design key Layer 2 functionalities, including forced transaction queues, safe blacklisting, and upgradeability. Through this analysis, we identify pitfalls in existing designs and introduce an enhanced model that has been model-checked to be correct. Finally, we propose a complete end-to-end methodology to analyze rollups' security and censorship resistance based on manually translating Alloy properties to property-based testing invariants, setting new standards.
\end{abstract}

\ccsdesc[500]{Security and privacy~Formal security models}
\ccsdesc[300]{Security and privacy~Security requirements}
\ccsdesc[300]{Security and privacy~Domain-specific security and privacy architectures}

\keywords{Blockchain Security; Scalability; Zero-Knowledge Proofs}


\maketitle

\section{Introduction}
\label{sec:introduction}

Blockchain technology, exemplified by major chains such as Bitcoin~\cite{nakamoto2008bitcoin} and Ethereum~\cite{wood2014ethereum}, has revolutionized finance and various other fields by enabling decentralized transactions. Despite its innovations, adopting these technologies has highlighted significant scalability challenges. The limited transaction throughput of these networks has led to ongoing research and development of various scalability solutions within the community~\cite{zhou2020solutions}.

Two primary strategies have emerged to enhance scalability. The first involves developing new blockchain architectures from scratch, designed for higher transaction throughput than traditional platforms like Ethereum~\cite{yakovenko2018solana,blackshear2023sui}, often at the expense of reduced security and network effects. The second strategy focuses on Layer 2 (L2) solutions, particularly rollups, which have become prominent in practise~\cite{thibault2022blockchain}. Rollups enhance scalability by executing transactions on a secondary blockchain (L2) and subsequently posting the state roots and transaction data back to the primary blockchain (L1). This approach allows rollups to inherit the security properties of the underlying L1 through the use of validity proofs~\cite{WhiteHat_roll_up_token}, i.e., Zero-Knowledge Proofs~(ZKPs)~\cite{goldwasser1985knowledge}, or fraud proofs~\cite{kalodner2018arbitrum}.

However, submitting transaction batches for verification by L1 contracts does not fully secure L2 users. Centralized control of L1 contracts through mechanisms like multisignature wallets poses some significant security risks. If a malicious party gains control of the majority of the keys, they could potentially redirect or steal funds. Notably, even if a proper governance protocol is in place, the issues that might occur from instant upgrades still hold. Further, the potential for censorship by L2 operators poses another major challenge, as users could be prevented from executing transactions. 

\subsection{Motivating Examples}
\label{sec:motivating}
\noindent
The compromise of multisigs has historically caused the greatest loss of funds in blockchains~\cite{zhou2023sok}. This was most recently demonstrated by the sophisticated exploit of ByBit's multisig,\footnote{\url{https://www.elliptic.co/blog/bybit-hack-largest-in-history}} resulting in the largest attack in blockchain history, with over~\$1~billion lost. No multisig setup can be considered fully secure. Critically, rollups today depend almost entirely on multisigs to operate their core L1 contracts, often without enforced or sufficient safety mechanisms.

As a result, current rollups do \textbf{not} truly inherit the security guarantees of their underlying L1s; instead, they depend on a small set of multisig keys, making them vulnerable to attacks that threaten their long-term viability. Centralized control over rollups further introduces serious censorship risks, as recent incidents illustrate.

In this work, we take up the challenge of formally specifying the mechanisms rollups must implement to inherit both the \emph{security} and \emph{censorship resistance} of L1. 
The following incidents highlight the urgent need for robust forced transaction queues, secure upgradeability mechanisms, and carefully controlled blacklist policies.
\point{Blast incident}
A recent incident involving the Blast rollup\footnote{\url{https://x.com/miszke_eth/status/1775196752993255613}}, which at the time held over~\$3 billion in total value locked~(TVL), highlights these issues. 
Following a major exploit in one of the protocols deployed on Blast, the operators quickly intervened to censor the attacker by upgrading the system. Although well-intentioned, this action highlighted that L2s currently do not offer the same level of security and censorship resistance as their underlying L1, as rapid changes could lead to substantial asset losses. Further, it highlights the upgrade power of the rollup operator without any safety mechanism in place to protect users. 
The code change\footnote{\url{https://etherscan.io/address/0xA280aEBF81c917DbD2aA1b39f979dfECEc9e4391\#code\#F1\#L449}} shown in Figure~\ref{fig:diff} highlights the rapidly deployed contract changes used by the Blast rollup operator to censor the attacker.

\begin{figure}[bt]
\begin{lstlisting}[language=Java, style=mystyle, basicstyle=\scriptsize\ttfamily,firstnumber=439]
address from = msg.sender;
if (msg.sender != tx.origin) {
  from = AddressAliasHelper.applyL1ToL2Alias(msg.sender);
}
// ...
require(
 from != 0x6E8836F050A315611208A5CD7e228701563D09c5 &&
 from != 0xc207Fa4b17cA710BA53F06fEFF56ca9d315915B7 &&
 from != 0xbf9ad762DBaE603BC8FC79DFD3Fb26f2b9740E87
);
\end{lstlisting}
\caption{Censoring changes in the Blast \texttt{OptimismPortal.sol}.}
\label{fig:diff}
\end{figure}

\point{dydx concerns}
Concerns regarding the excessive authority of the rollup operator to execute a rug pull~\cite{sun2024sok} on its users apply to the dydx rollup, which is controlled via a~3-out-of-5 multisig, which means that only three rogue parties are enough to hijack the entire rollup and lead to an immediate code upgrade. Citing the postmortem,~\footnote{\url{https://www.odaily.news/en/post/5191266}} this issue could lead to draining the rollup.
\begin{shadedquotation}
\scriptsize
Anyone who wants to upgrade the Layer 2 contract must be subject to the time lock delay limit, and the contract upgrade should take effect later than the mandatory withdrawal. For example, the contract upgrade of dydx now has a delay of at least~48 hours, so the delay for the forced withdrawal/escape hatch mode to take effect should be reduced to within~48 hours. In this way, after users discover that the dYdX project team wants to incorporate malicious code into the new version of the contract, they can withdraw their assets from Layer~2 to Layer~1 before the contract is updated.
\end{shadedquotation}
In the rest of this paper, we show that the situation is even more complex: as demonstrated in Section~\ref{sec:upgradeability}, a timelock alone is insufficient; a secure upgrade process that the rollup operator must follow, regardless of who controls the upgrade decision, is essential.

\point{Linea hack}
In~June 2024, the Linea chain was halted in response to a project on Linea that was compromised, and some addresses were black-listed. Citing the Linea team's communications\footnote{\url{https://x.com/LineaBuild/status/1797283402745573837}}\footnote{\url{https://cryptoslate.com/linea-under-scrutiny-for-unilateral-block-production-halt-amid-velocore-hack/}}; this issue had similar implications to the \emph{Blast incident}:
\begin{shadedquotation}
\scriptsize
The sequencer was paused from block~5081800 and~5081801. During this pause, we gave the @Velocorexyz time team to support their efforts of triaging the vulnerability. We also censored the hacker's addresses. This significantly reduced the ecosystem impact on Linea users.
...
Linea's team made a decision to halt block production by pausing the sequencer and censor attacker addresses to protect the users and builders in our ecosystem. Like other L2s, we are still in the "training wheels" phase of existence, giving us safeguards to use.
...
Meanwhile, teams at Velocore and Linea have requested to CEX to freeze the exploited funds, and Velocore is setting up an onchain negotiation process.
\end{shadedquotation}

\subsection{Paper Overview}
In this work, we are the first, to the best of our knowledge, to formalize and analyze essential mechanisms that L2 rollups must implement to truly inherit the security and censorship resistance of their underlying L1 blockchains. Using the Alloy specification language, we define and test critical security properties these mechanisms should exhibit. Our framework aims to guide developing and testing robust L2 mechanisms, providing concrete security guarantees to L2 users. We focus on ZK-Rollups due to their simple L1 logic, and explore the following critical mechanisms:

\point{Forced Queue} A mechanism implemented in L1 contracts allowing users to bypass potential L2 censorship by submitting transactions directly to the L1. This mechanism is supported, or planned to be supported, by some of the top rollups, both optimistic and ZK-Rollups, as shown on the L2beat Risk page.\footnote{\url{https://l2beat.com/scaling/risk}}

\point{Blacklisting} Recognizing that strong guarantees of the previous mechanism could lead to regulatory issues, we propose a secure mechanism for enforcing blacklisting policies. This mechanism prevents immediate blocking of users, as highlighted in Section~\ref{sec:motivating}, and should ideally be updated through a governance process. 

\point{Upgradeability} We outline a secure approach to upgrading system components that preserves user security and is compatible with previous mechanisms. Recent rollup incidents further justify the need for secure upgrades. 
We also highlight common pitfalls in the state-of-the-art upgradeability mechanisms and demonstrate potential vulnerabilities through counter-example-driven reasoning. 

In this work, we introduce mechanisms grounded in formal modeling of rollup protocols designed to mitigate issues such as those discussed in Section~\ref{sec:motivating}. Specifically, we describe a safe upgradeability mechanism that is required to resolve all three examples (Section~\ref{sec:upgradeability}). Additionally, for scenarios outlined in the first and third examples, we describe a secure blacklisting mechanism (Section~\ref{sec:blacklisting}) that further helps address the issues.

\subsection{Contributions}
\begin{itemize}
    \item \textbf{Formal model for ZK-Rollups:} Together with a threat model (Section~\ref{sec:threat-model}), we introduce the first formal model for ZK Rollup smart contracts operating on Layer~1 blockchains, responsible for the security of L2 (Section~\ref{sec:strawman}). This model is designed to be both adaptable and extensible, facilitating comprehensive analyses of core design properties essential for the development of secure ZK-Rollups. This model aims to guide implementers and operators of ZK-Rollups. 

    \item \textbf{Analysis of key mechanisms:} We conduct a detailed analysis of three essential mechanisms integral to ZK-Rollups: the forced queue, safe blacklisting, and upgradeability (Section~\ref{sec:alloy}). Our analysis identifies common design pitfalls and offers robust designs that integrate these mechanisms securely, ensuring they function correctly. We also outline security properties that each mechanism must satisfy. We aim to provide bounded model-checked models for these mechanisms as opposed to post-factum analysis of either code or designs.

    \item \textbf{Evaluation of state-of-the-art mechanisms:} Using our Alloy model, we analyze current ZK Rollup designs, highlighting issues in the \emph{design} of upgrade policies of one rollup with concrete counterexamples (Section~\ref{sec:testing}). 
    We show that our proposed upgradeability mechanism is secure, setting a new standard for ZK Rollup security. 
    We also discuss how the Alloy models can be used to guide the verification/testing of the implementation of the proposed mechanisms and discuss the instantiation of this methodology to test Scroll's implementation (Section~\ref{sec:endtoend}).
\end{itemize}
We share our Alloy code at the following location: \url{github.com/StefanosChaliasos/zk-rollup-security}, and the implementation of property-based testing for Scroll at: \url{https://github.com/StefanosChaliasos/scroll-contracts/}.

\point{Note on analyzing optimistic rollups}
In this work, we focus on ZK-Rollups rather than optimistic rollups. Extending our formal model to optimistic rollups would require accurately modeling the fraud-proof mechanisms, including challenge periods, dispute resolution games, and fault proofs, significantly complicating the on-chain logic. These mechanisms are complex to model formally and are beyond the scope of this work that focuses on the mechanisms required to be implemented by rollups to inherit the security and censorship resistance of the L1. For this reason, we leave the formal modeling and analysis of optimistic rollups as an important direction for future work.

\section{Background}
\label{sec:background}

\subsection{Blockchain Scalability and ZK-Rollups}

Blockchain scalability has been a persistent challenge, particularly for established networks like Ethereum~\cite{wood2014ethereum}, which processes only tens of transactions per second~(TPS).\footnote{\url{https://l2beat.com/scaling/summary}} The scalability trilemma~\cite{monte2020scaling} posits that blockchain systems cannot simultaneously achieve scalability, decentralization, and security without compromises. This has led to efforts to enhance scalability, focusing on two primary strategies: base layer scaling (L1) and Layer 2 (L2) scaling solutions. Base layer scaling, which includes techniques such as sharding~\cite{wang2019sok} and novel consensus protocols~\cite{lashkari2021comprehensive}, involves either the modification of existing blockchains -- a complex and daunting task -- or the development of new blockchain architectures. While modern blockchains like Solana~\cite{yakovenko2018solana} and Sui~\cite{blackshear2023sui} have demonstrated success, they often lack the established security, liquidity, and comprehensive ecosystem found in legacy blockchains like Ethereum.\footnote{As of 14/4/2025, Ethereum has~51.66\% of the total TVL for all chains, according to~\url{https://defillama.com/chains}.}


Rollups~\cite{thibault2022blockchain} have emerged as hybrid L2 solutions, distinguishing themselves by offloading computation off-chain while retaining and validating data on-chain, thus addressing the data availability issue while inheriting L1's security. Rollups batch and execute transactions on an L2 blockchain and submit the results into L1.

The execution of transactions in the L2 instead of L1 allows rollups to process significantly more transactions per second than their L1 counterparts. By submitting a single L1 transaction for a batch, which might include up to 4K L2 transactions at the moment\footnote{That is the current batch size for zkEsync Era.}, to the underlying blockchain, rollups not only scale the total TPS but also ensure data availability and inherit the security properties of the L1 network. Altering the L2 state recorded on L1 would require breaking substantial security, which would be both difficult and costly. This architecture enables rollups to offer an efficient and secure scaling solution for legacy blockchains. 
This model, i.e., rollup-centric scaling, has gained traction as the \textit{principal} method for scaling Ethereum, with two predominant variants: optimistic rollups~\cite{kalodner2018arbitrum} and ZK-Rollups~\cite{WhiteHat_roll_up_token}.\footnote{As of~14/4/2025, rollups have more than~33B USD TVL according to \url{https://l2beat.com}.}

Optimistic rollups rely on a system of trust, assuming transactions are valid unless challenged, which makes the implementation of the L2 simpler but introduces delays and adds complexity to rollup's contracts logic due to the fraud-proof mechanism. In contrast, ZK-Rollups use ZKPs to validate transactions in the L1 as soon as a proof has been created off-chain and submitted to the L1 contracts, ensuring faster finality but a higher computational and complexity cost in the L2 level. While Optimistic rollups are generally easier to implement, they suffer from potential delays in withdrawals (currently, most L2s apply a~7-day long challenge period); ZK-Rollups, on the other hand, offer faster finality and more compression opportunities (e.g., state diffs~---~not all transaction data need to be posted in the L1) to the data submitted to the L1, leading to smaller costs.

\begin{figure}[t]
    \centering
    \includegraphics[width=0.36\textwidth]{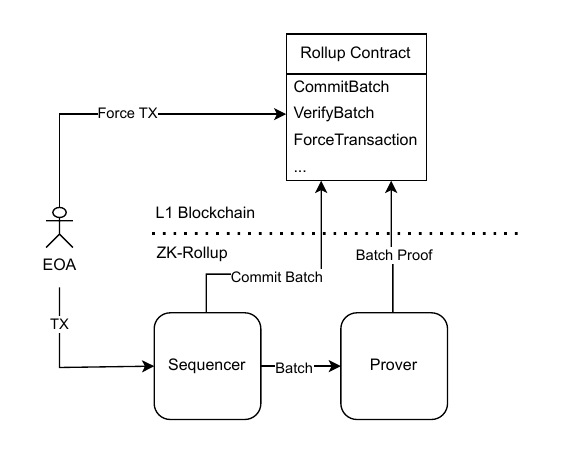}
    \caption{High-level architecture overview of ZK-Rollups. EOA: Externally owned address.}
    \label{fig:zk_rollup_architecture}
\end{figure}

\point{Components of ZK-Rollups}
As illustrated in Figure~\ref{fig:zk_rollup_architecture}, ZK-Rollups consist of several key L2 components, including the \textit{Sequencer} and the \textit{Prover}. The Sequencer is responsible for processing, ordering, and executing transactions into batches. Although currently often centralized, there is ongoing research aimed at decentralizing this component to enhance system trustworthiness and resilience~\cite{motepalli2023sok}. The Prover, on the other hand, generates cryptographic proofs of the correctness of these batches. Note that both components can be split into various other components (e.g., executor, L1-L2 relayer, and prover coordinator). The rollup's smart contracts, deployed on the L1, are essential to the overall architecture, executing various critical operations. These contracts must be analyzed to determine if they meet the properties detailed in Section~\ref{sec:threat-model}.

\point{L1 Smart Contract Operations} L1 contracts are used to \emph{(i)} \textit{commit the transaction batches} prepared by the Sequencer, which includes posting transaction data necessary for proof generation, verification, and reproduction of the L2 state. L1 contracts also \emph{(ii)} \textit{verify the cryptographic proofs} produced by the Prover to ensure that all transactions in a batch are correctly executed. Further, \emph{(iii)} in cases where transactions need to be pushed through despite potential censorship, downtime, or adversarial actions by the L2 operators, the L1 contracts can \textit{enforce transactions}. Smart contracts on L1 also \emph{(iv)} manage \textit{protocol upgrades}, which can include changes to the rollup's operational logic or security parameters, ensuring that upgrades occur smoothly without compromising ongoing operations or security. Finally, \emph{(v)} these contracts manage \textit{cross-layer transactions} for depositing and withdrawing from the L2.

\point{Transaction lifecycle in ZK-Rollups} The process begins when a user signs and submits a transaction to the L2 network, marking it as \textit{Pending}. It quickly becomes \textit{Preconfirmed} once the sequencer processes and includes the transaction in a block. 
The transaction then becomes \textit{Committed} when included in a batch that is submitted to the L1 contract, which allows for reconstruction and verification of the L2 state based on data posted to L1. Finally, the transaction reaches the \textit{Verified/Finalized} stage when the batch's proof is validated by the L1 contract, ensuring the transaction and its associated state are immutable and secure on L2. Additionally, transactions can also be submitted directly through the L1 contracts in scenarios where L2 operators may be censoring transactions or in cases of L2 downtime, ensuring robustness against operational failures or malicious activities.

\subsection{Formal Modeling of Software Design}

Formal modeling enhances the correctness and robustness of software systems through a rigorous analysis of their designs. By providing a precise mathematical description, formal modeling facilitates the verification of properties and behaviors before implementation, thereby reducing errors and improving system reliability.

Alloy~\cite{jackson2002alloy, jackson2012software} is a declarative specification language using first-order relational logic to model complex systems succinctly, ideal for expressing constraints and behaviors in software designs. The Alloy Analyzer automates checking these models against specified properties, leveraging SAT solvers to find solutions or identify inconsistencies within a bounded scope.

Alloy has been used in various research projects to model and verify the properties of software architectures and protocols. For instance, Akhawe et al.~\cite{akhawe2010towards} utilized Alloy to model web security mechanisms, demonstrating the language's utility in capturing complex interactions and verifying security properties while revealing three previously unknown vulnerabilities. This exemplifies how Alloy facilitates a deep understanding and validation of complex systems in practical applications.

In this paper, we use Alloy 6~\cite{brunel2021alloy6} to specify and examine the security attributes of ZK-Rollups. We create formal models of the contract interactions within ZK-Rollups, outlining key properties for mechanisms such as forced transactions and upgradeability. Using the Alloy Analyzer, we not only validate the accuracy of certain designs, but also uncover weaknesses in simplistic designs, highlighting their security issues. 
To quote the Alloy book~\cite{jackson2012software}, \emph{``In this respect, the Alloy language and its analysis are a Trojan horse: an
attempt to capture the attention of software developers, who are mired in the tar pit of implementation technologies, and to bring them back to thinking deeply about underlying concepts''}.



\section{Threat Model and Goals}
\label{sec:threat-model}

\subsection{Threat Model}

We construct a robust adversary model to rigorously assess the security of the mechanisms presented in this paper. This model assumes that the L2 network is malicious, with no trust assumptions regarding its correct operation. We consider an adversary capable of gaining complete control over the L2 network, emphasizing the importance of properly designing ZK-Rollups' mechanisms.

\point{Principals} The main principals in ZK-Rollups are \textit{users}, \textit{rollup operators}, and \textit{rollup administrators}.

\point{Users}
Users interact with the L2 either by submitting their transactions directly to the L2 network, or by submitting them through the rollup's contract on the L1 via the forced queue.

\point{Rollup operators}
Rollup operators handle the components that sequence user transactions in the L2 into batches, commit them in the L1, and request validity proofs (c.f.\ Figure~\ref{fig:zk_rollup_architecture}). They may also receive the proofs and submit them to the L1. Currently, in most L2s, there is a single centralized entity operating the L2, but in our model, the operator could be controlled by one or many entities that do not need to be trustworthy.

\point{Rollup administrators}
In addition to rollup operators, some entities have administrator capabilities for the L1 smart contracts of the L2. 
Administrators can be white-listed or elected via governance mechanisms. 
These administrators have the ability to request system upgrades or update blacklists. In practice, the administrator capabilities are assigned to a multisig that could be controlled by some distinct entities. The mechanisms we describe in this paper should maintain security even if administrators are malicious. 

\point{Security assumptions}
\begin{itemize}
    \item The underlying \emph{L1 remains secure} and maintains liveness to enable progress of L2 operations. Should L1 face downtime, L2 operations might be delayed as well, but the security of the L2 would remain intact, thus preserving system safety.

    \item The \emph{cryptographic primitives (ZKPs) and their implementation are assumed to be secure}~\cite{chaliasos2024sok}, ensuring that no verifier accepts an invalid proof of execution. This assumption also extends to the correctness of any trusted setup procedures, the deployed verification keys on the L1 verifier contract, and the smart contract code responsible for proof verification.

    \item \emph{Smart contracts operating L2 functionalities on the L1 are considered secure} against \emph{conventional} vulnerabilities, such as re-entrancy attacks~\cite{chaliasos2024smart}, and are modifiable only through a well-defined upgrade process. In other words, in this paper, we focus on what may be called logical vulnerabilities. 
    
    \item \emph{The initial state of the system} is empty, i.e., the system has not processed or finalized any transactions and computed any proofs. Furthermore, the system's initial policies and contracts are non-malicious and secure for users, meaning that if a user can interact with the L2, the user's funds cannot be locked or lost before an adversarial upgrade is enforced.
\end{itemize}

\point{Strong Adversary Model}
Any of the parties mentioned above~---~users, operators, and administrators~---~can become malicious. Users can perform DoS attacks, operators can engage in transaction censorship, and rollup administrators can execute malicious upgrades that could lead to the theft of all user assets. 
Despite these possibilities, the saving grace is that rollup contracts deployed on the underlying L1, together with the economic security that the L1 provides, allow us to use decentralized trust to address these issues. This comes down to the ability to analyze~\cite{ouroboros} these rollup contracts. 

For example, zkSync Era has contracts such as \texttt{Verifier.sol}, which process validity proofs and can be upgraded by a multisig\footnote{\url{https://github.com/matter-labs/era-contracts}}. It is critical that this verifier can only be updated through a safe process because if it becomes malicious, users can lose their money. Hence, users should have a sufficient mechanism to exit the rollup in case of adversarial or questionable upgrades. 
While our model is extensible, this paper largely focuses on functional security, sidestepping DoS concerns. 

\subsection{User-centric Security Goals}
In this work, our objective is to provide robust safety guarantees to L2 users, even under a strong adversarial model where the L2 network and L2 administrator can be fully compromised.

\point{Weak liveness}
A key goal is to ensure that any user transaction submitted to the forced queue is processed as long as the L2 continues to make progress. This guarantee is crucial to protecting users against potentially compromised or malicious L2s or even adding L1-like censorship resistance for L2s. Moreover, our proposed design ensures that even when L2 contracts enforce transaction blacklisting, transactions cannot be halted if they have already been submitted to the forced queue. This mechanism is vital to maintaining the integrity and safety of ZK-Rollups. In summary, weak liveness means that the L2 will eventually process user transactions submitted through L1 (i.e., inheriting the liveness of the L1), or the rollup will be frozen. In Section~\ref{sec:forcedqueue}, we design a forced queue mechanism that enforces weak liveness.
\point{Secure upgrades and user assets protection}
In addition, we provide and analyze designs for securely upgrading L2's contracts. In practical terms, this means that if there's an undesirable upgrade, users have time to exit the rollup. This includes mechanisms to prevent ``rug pulls'' (see Section~\ref{sec:motivating} for some examples of this) or sudden, unanticipated changes that may compromise user assets. Ensuring that users receive sufficient notice and have effective mechanisms to respond to changes is essential for maintaining trust and stability within L2 environments.

\section{Implementation in Alloy}
\label{sec:alloy}
In this section, we present the implementation details of our ZK Rollup model using the Alloy specification language. We first give a brief technical introduction to Alloy, followed by the Alloy model for a \emph{strawman} ZK Rollup and three critical mechanisms, highlighting our reasoning and essential security properties. In Section~\ref{sec:testing}, we demonstrate how to use our model to detect issues in flawed designs. Our approach directly encourages finding problems early, at the design stage, as opposed to later.

\point{Why Alloy}
We selected Alloy 6~\cite{brunel2021alloy6} over other formal methods tools, such as TLA+ and NuSMV, due to its balance of expressiveness, \emph{ease of specification}, and automation for finding counterexamples. Alloy's declarative syntax based on first-order relational logic is particularly well-suited for modeling relational systems, such as rollup states and transitions~\cite{jackson2002alloy}. Prior work in web security~\cite{akhawe2010towards} and smart contract validation~\cite{godoy2022predicate} shows that Alloy enables uncovering subtle vulnerabilities effectively, making it a practical and proven choice for the security analysis of systems where quick design iteration and counterexample-driven debugging are critical.

\subsection{A Brief Introduction to Alloy}
Alloy 6~\cite{jackson2002alloy,brunel2021alloy6} is a declarative modeling language that uses \textit{first-order relational logic} to specify and check the properties of software systems. Data types in Alloy are represented as relations defined by \emph{signatures}, which can be abstract or concrete. \emph{Abstract signatures} categorize general types that are refined by more specific \emph{concrete signatures}, creating a flexible and powerful type system.
The language enforces the integrity of the model through \emph{facts}, which are constraints that must always hold within the model. \emph{Functions} and \emph{predicates} in Alloy are named expressions and logical formulas, respectively; these allow the parametrization and reuse of complex logical constructs throughout the model.

Reasoning about models in Alloy is facilitated by the \emph{Alloy Analyzer}, which translates the code into a satisfiability problem. The resulting SAT instance is then tackled using SAT solvers, which amounts to \emph{bounded model checking} within a specified model \href{https://alloy.readthedocs.io/en/latest/language/commands.html\#scopes}{\emph{scope}}. Commands \texttt{check} and \texttt{run} are instrumental in this process: \texttt{check} is used to verify whether a given predicate can be invalidated, thus checking for counterexamples within the constraints of the model, while \texttt{run} finds instances that satisfy all conditions of the model up to a certain scope, allowing effective testing of various scenarios.

The Alloy analysis is always limited to a specified scope. The \emph{small-scope hypothesis} states
that a high proportion of errors can be found by analyzing a
system for all test inputs within some small scope~\cite{jackson1996elements,jackson2019alloy}.
However, Alloy's guarantees are strictly bounded: if no counter-example appears within the specified scope, nothing can be inferred beyond it. To offset this incompleteness, we rely on the small-scope hypothesis. Still, bounded success is \emph{not} a universal proof. Alloy is a performant bug-finder, not a full verifier. We discuss this trade-off further in Section~\ref{sec:discussion}.

This structured approach to modeling with Alloy ensures model checking of properties and behaviors in software systems, making it an invaluable tool in the development and analysis of reliable software architectures. In the upcoming sections, we present how we used Alloy to design and verify the correctness of critical elements of ZK-Rollup logic.

\subsection{Strawman ZK Rollup Model}
\label{sec:strawman}

In this section, we formalize a simple ZK rollup model that notably lacks mechanisms to process forced transactions through its L1 contract and similarly lacks any mechanism for safely upgrading its contracts. 
In this model, rollup operators and administrators may arbitrarily censor user transactions. Furthermore, in scenarios where the L2 becomes malicious or experiences downtime, user funds can be frozen. In subsequent sections, we will enhance this model to include support for forced transactions and enable safe upgrade protocols.

\point{Data model specification}
The data model of our simple ZK rollup is structured around the following key elements:

\begin{itemize}[leftmargin=10pt]\itemsep=0pt
\item \textbf{Inputs:} Represented by abstract datatypes, inputs represent transactions submitted by end-users. This model deliberately omits additional attributes such as sender, receiver, or transaction amounts, as these details are not particularly relevant to our analysis.

\item \textbf{Block:} Blocks hold an ordered sequence of inputs, forming the basic building block of our model. Although our model could be extended to include batches of blocks, we currently limit it to blocks for simplicity.

\item \textbf{Commitment and proof:} We track two fields for both commitments and proofs: \emph{state} and \emph{diff}. The state is represented as an ordered sequence of blocks, and the diff refers to the current block for which we commit to the L1 contract and produce a proof. It is important to note that in practice, the diff could encompass all transaction data for the block or even a state diff.

\item \textbf{L1:} This component models the L2 rollup's representation on the underlying L1 chain, incorporating the finalized state of blocks along using sets of submitted commitments and proofs.
    
\end{itemize}    
In typical rollup systems, the state could be represented by a short digest (for example, the root of the Merkle tree), but in our model, we represent a state by the sequence of all finalized blocks from the genesis block.
The Alloy code describes the data model as follows. Note that \href{https://alloy.readthedocs.io/en/latest/language/signatures.html#seq}{\texttt{seq}} indicates an ordered list in Alloy and \href{https://alloy.readthedocs.io/en/latest/language/signatures.html#one}{\texttt{one}} is a multiplicity which forces variable to always be a singleton set. 
\begin{lstlisting}[language=alloy,style=mystyle]  
var sig Input{}
var sig Block {  var block_inputs : seq Input }

var abstract sig ZKObject {
  var state : seq Block,
  var diff : one Block 
}
var sig Proof extends ZKObject{}{
  not state.hasDups
  diff not in state.elems
}
var sig Commitment extends ZKObject{}{
  not state.hasDups
  diff not in state.elems
}
one sig L1 {
  var finalized_state : seq Block,
  var commitments : set Commitment,
  var proofs : set Proof,
}{ not finalized_state.hasDups }
\end{lstlisting}

\point{Processing logic and events}
The core functionalities of the L1 contract(s) are represented by a series of actions that are essential to the operation of the system; these actions are:
\begin{itemize}[leftmargin=10pt]\itemsep=0pt
\item\textbf{Receiving commitments:} The L1 contract is responsible for receiving and storing commitments. It ensures there is no duplication (line~2) by checking that the commitment does not already exist within the current set of commitments. Additionally, it maintains a sequence that extends the existing state of the L2 (lines~3 and~4). This is achieved by verifying that the state sequence within the commitment aligns precisely with the current finalized state up to the last element. This check ensures that each new commitment builds directly upon the most recent confirmed state, thereby preserving the integrity and continuity of the blockchain's transaction history.

\begin{lstlisting}[language=alloy, style=mystyle]  
pred receive_commitment[c : Commitment] {
  no c & L1.commitments 
  c.state.subseq[0,sub[#L1.finalized_state,1]] = L1.finalized_state 
  L1.commitments' = L1.commitments + c
}
\end{lstlisting}
Alloy supports common set operators like membership (\verb;in;), intersection (\verb;&;), union (\verb;+;), etc.
To refer to an expression in the next state we use the  apostrophe operator \verb;';. For example, \verb;x' = x + c; means that the value of \verb;x; in the next state is the union of \verb;c; and the value of \verb;x; in the current state.

\item\textbf{Receiving proofs:} Similar to commitments, proofs are also received and stored by the L1 contract. These proofs are aligned with the current state to maintain consistency and ensure the validity of the L2 as in the case above. 
    
\item\textbf{Single-step rollup update:} This critical functionality of the L1 contract involves processing both commitments and proofs to securely advance the state of L2, as outlined in the Alloy predicate \texttt{rollup\_process}. The process begins by confirming that both the commitment \( c \) and the proof \( p \) are present within the respective sets managed by the L1 contract (lines~2--3). It then verifies that both the commitment and the proof agree on the current state and the proposed state transition (lines~4--6), affirming that they are intended to advance the L2 from the same state. The predicate also ensures that the proposed state transition, or diff, has not been previously processed (line~7, \verb;s.idxOf[x]; returns the first index where \verb;x; appears in \verb;s;). After this check, the finalized state of the L2 is advanced to include this diff (line~8); commitment and proof sets are updated to remove processed or outdated entries (lines~9--10), maintaining the rollup's integrity.

\begin{lstlisting}[language=alloy,style=mystyle]  
pred rollup_process[c: Commitment, p:Proof] {
    c in L1.commitments  
    p in L1.proofs
    c.state = p.state
    c.diff  = p.diff
    c.state = L1.finalized_state
    (no L1.finalized_state.idxOf[c.diff])
    L1.finalized_state' = L1.finalized_state.add[p.diff]
    L1.proofs' = L1.proofs - (p + { q : Proof | #q.state < #L1.finalized_state })
    L1.commitments' = L1.commitments - (c + { q : Commitment | #q.state < #L1.finalized_state })
}
\end{lstlisting}

\end{itemize}
While these are omitted for brevity, our Alloy implementation includes \emph{frame conditions} to ensure that parts of the system that are not affected by a particular operation remain unchanged. 

\point{Strawman rollup properties (SRPs)}
Transaction processing within the rollup is implemented as a series of events, obeying the following rules.
\srp{Event Granularity} At any given time, only one event occurs. This property ensures that the granularity of events is correct and that we do not have overlapping events in the model of the system.
\srp{Monotonic State} The finalized state is guaranteed to grow monotonically. In other words, at any given moment, the system can only append new blocks to the finalized state, but previously finalized blocks never change.
\begin{lstlisting}[language=alloy,frame=lines,basicstyle=\ttfamily\footnotesize,columns=fixed]  
always (finalized_state in finalized_state')
\end{lstlisting}
\srp{Justified State} If a block reaches the finalized state, then \href{https://alloy.readthedocs.io/en/latest/language/time.html}{once} (i.e., at least at one point in the history) there was a proof and a commitment for it that corresponded to the then-current state, ensuring all finalized transactions are valid and verified.
\begin{lstlisting}[language=alloy,style=mystyle]  
always(
 all b : Block | some L1.finalized_state.idxOf[b]
   implies
    (once some c : Commitment, p : Proof |
     c in L1.commitments
     and p in L1.proofs and 
     c.diff = p.diff and b = c.diff
     and c.state = L1.finalized_state)
)
\end{lstlisting}
\srp{State Progression Validity} A commitment or proof that is smaller than the current state is never successfully processed.

\point{Checking} As we mentioned above, the Alloy Analyzer does bounded model checking; the scope of checking is provided. 
For example, the following command checks whether property \emph{SRP2} holds in the scope, which contains~10 objects of every signature and in traces up to 20 steps.
\begin{lstlisting}[language=alloy,frame=lines,basicstyle=\ttfamily\footnotesize,columns=fixed]  
check monotonic_state for 10 but 1..20 steps
\end{lstlisting}
The \href{https://alloy.readthedocs.io/en/latest/language/commands.html#run}{\texttt{check}} command tells the Alloy Analyzer to find a counterexample within the specified scope. If the counterexample is not found, we know that the property holds within the specified scope.

Strictly speaking when we checked the property for 20 steps it still can fail for 21 steps. However, \emph{small-scope hypothesis}
states that a high proportion of errors can be found by checking a
program for all test inputs within some small scope~\cite{jackson1996elements}.
 
\subsection{Forced Queue}
\label{sec:forcedqueue}
In the previous section, we described a strawman ZK Rollup that does not support user transactions submitted through the L1 to force their inclusion by the L2. This section extends the strawman model to enable that functionality, providing users with stronger guarantees. Our goal is to ensure that forced transactions are processed by the L2; otherwise, the L2 cannot progress (i.e., extend the finalized state). 

\point{Data model extensions}
We introduce a new abstract signature called \texttt{ForcedEvent}, and another signature called \texttt{ForcedInput} that extends it. \texttt{ForcedInput} contains a field \texttt{tx} that represents a transaction submitted to the contracts of the L2 in the L1. Additionally, in the \texttt{L1} type, we add a field called \texttt{forced\_queue} that is an ordered sequence of \texttt{ForcedEvents}. The relevant Alloy code is provided below.

\begin{lstlisting}[language=alloy,style=mystyle]
var abstract sig ForcedEvent {}

var sig ForcedInput extends ForcedEvent {
   var tx : one Input
}

one sig L1 {
    var forced_queue : seq ForcedEvent
    // + previously introduced fields
}
\end{lstlisting}

\point{Processing logic}
Initially, we define the logic to add forced inputs to the forced queue. This process simply checks that the forced input does not already exist in the forced queue and adds the new input to the end of the queue. The respective Alloy predicate is as follows:

\begin{lstlisting}[language=alloy,style=mystyle]
pred receive_forced[f : ForcedEvent] {
 no L1.forced_queue.idxOf[f]
 L1.forced_queue' = L1.forced_queue.add[f]
}
\end{lstlisting}
Next, we update the \texttt{rollup\_process} predicate to take into account the forced queue. If the forced queue is empty, then the rollup should process transactions as before. If it is not empty, then every new finalized block needs to include the head of the forced queue and can also include other transactions from the forced queue.

\begin{lstlisting}[language=alloy,style=mystyle]
some L1.forced_queue
   implies 
    (L1.forced_queue.first in ForcedInput 
     and some c.diff.block_inputs.idxOf[L1.forced_queue.first.tx])
\end{lstlisting}
Upon being finalized, elements need to be removed from the forced queue. In Alloy, we ensure that no transactions in the current processing batch remain in the updated forced queue (line~1). The FIFO transaction order is maintained; transactions are moved closer to the queue's head to expedite processing while preserving their relative positions (line~3). This preservation of order is crucial for fairness and for ensuring that, eventually, all transactions are processed. We also check that the relative positions of the elements do not change (line~8). Finally, to maintain the integrity of the forced queue, the model prohibits adding new elements during processing (line~12).

\begin{lstlisting}[language=alloy,style=mystyle]
no (L1.forced_queue'.elems.tx & p.diff.block_inputs.elems)

all x : ForcedInput | (x.tx not in p.diff.block_inputs.elems  
    and (some L1.forced_queue.idxOf[x]))
   implies L1.forced_queue'.idxOf[x] < L1.forced_queue.idxOf[x]
    and (some L1.forced_queue'.idxOf[x])

all x, y : ForcedEvent | some L1.forced_queue'.idxOf[x] and some L1.forced_queue'.idxOf[y]
    and L1.forced_queue.idxOf[x] < L1.forced_queue.idxOf[y] implies
    L1.forced_queue'.idxOf[x] < L1.forced_queue'.idxOf[y]

all x : ForcedEvent | x not in L1.forced_queue.elems
   implies x not in L1.forced_queue'.elems
\end{lstlisting}

\point{Forced queue properties (FQPs)}
We summarize the desired properties of the forced queue below.

\fqp{Guaranteed Processing} If the forced queue is non-empty and a new state is finalized, then the head of the forced queue must be processed and removed.
\begin{lstlisting}[language=alloy,style=mystyle]
always (
 (some L1.forced_queue and some (L1.finalized_state' - L1.finalized_state))
   implies
     L1.forced_queue.first.tx in new_finalized_inputs
    and not L1.forced_queue.first.tx = L1.forced_queue'.first.tx 
)
\end{lstlisting}
\fqp{Forced Queue Stable} If the finalized state did not change, then the forced queue did not decrease.
\fqp{State Invariant} If the forced queue is non-empty and did not change, then the finalized state remains unchanged.
\fqp{Forced Inputs Progress} Forced inputs that were not processed move closer to the head of the forced queue.
\begin{lstlisting}[language=alloy,style=mystyle]
always (
 (some L1.forced_queue 
 and #L1.finalized_state < #L1.finalized_state') implies
 (all x : ForcedEvent |
   (some L1.forced_queue.idxOf[x]
   and some L1.forced_queue'.idxOf[x])
    implies 
     L1.forced_queue'.idxOf[x] < L1.forced_queue.idxOf[x])
)
\end{lstlisting}
\fqp{Order Preservation} Forced inputs retain their relative order within the queue.
\fqp{Finalization Confirmation} If an input was in the forced queue and then disappeared from it, it was finalized.
\begin{lstlisting}[language=alloy,style=mystyle]
always (
 all fi : ForcedInput | fi in L1.forced_queue.elems 
  implies always(fi not in L1.forced_queue.elems implies fi.tx in all_finalized_inputs)
)
\end{lstlisting}
These properties combined result in the following guarantee that ensures that a user's transactions can be processed, even if the L2 operators arbitrarily and immediately decide to censor that user.

\highlightbox{Guaranteed Transaction Processing in the Forced Queue}{
For any transaction \( t \) in the forced queue of an L2, if the L2 progresses, then \( t \) will eventually be processed. 
}

\subsection{Forced Queue with Blacklisting}
\label{sec:blacklisting}
Recently, transaction and address censorship have become prominent issues in the  Ethereum ecosystem. For example,~6\% of Ethereum blocks are constructed to be OFAC-compliant~\cite{wahrstatter2024blockchain}, and Circle~(USDC) maintains a list with blacklisted addresses~\cite{wang2023blockchain}. Some L2 networks may choose to censor transactions to be OFAC compliant, potentially to be determined through their governance protocol. Next, we describe an ``enshrined'' blacklisting mechanism; this mechanism was meticulously designed, formalized, and checked with Alloy Analyzer to ensure it does not violate forced queue properties. 

A naive blacklisting mechanism could violate the main property of the forced queue that a transaction in the forced queue must be processed, or it could be abused to prevent users from exiting the L2. Our mechanism receives blacklisting policies through the forced queue as special inputs that can be processed only when they are at the head of the queue. Importantly, once a blacklisting policy is processed, new transactions that do not comply with this policy cannot be added to the forced queue. Note that if a transaction that would otherwise be blacklisted by the policy is already in the queue before the policy is processed, then the L2 must process it; otherwise, the system will freeze.

\point{Data Model Extensions}
Two changes are required to support blacklisting. First, we introduce a new signature that extends \texttt{ForcedEvent} called \texttt{ForcedBlacklistPolicy}, which describes a set of inputs that should be rejected. In practice, this could target anything from specific addresses to specific transactions. Here, for simplicity, we narrow it down to a set of \texttt{Inputs}. Furthermore, we include a new field \texttt{blacklist} in \texttt{L1}, which is a set of \texttt{Inputs} that holds the currently active blacklisting policy.

\begin{lstlisting}[language=alloy,style=mystyle]
var sig ForcedBlacklistPolicy extends ForcedEvent {
 var predicate : set Input
}

one sig L1 { var blacklist : set Input }
\end{lstlisting}

\point{Processing Logic}
To integrate blacklisting policies into our forced queue mechanism, we need to make some changes to our model. Firstly, we update \texttt{receive\_forced} to append blacklist updates and only accept forced inputs that are not currently blacklisted by the active or pending forced blacklist policy.

\begin{lstlisting}[language=alloy,style=mystyle]
no L1.forced_queue.idxOf[f]
\end{lstlisting}
Next, we need to add a predicate to update the blacklist in the contract when it reaches the head of the forced queue. This predicate enforces that the blacklist is updated and the blacklist policy is removed from the queue.

\begin{lstlisting}[language=alloy,style=mystyle]
pred update_blacklist[f : ForcedBlacklistPolicy] {
  L1.forced_queue.first = f
  L1.blacklist' = L1.forced_queue.first.predicate
  L1.forced_queue' = L1.forced_queue.delete[0]
}
\end{lstlisting}
Finally, we update the predicate \texttt{rollup\_process} to prevent processing blacklisted blocks (line~1) and only allow processing inputs that are ahead of the next blacklist policy in the forced queue (line~3).
It is important to understand that blacklisting changes the operation of the forced queue. Now, if a new blacklisting policy was added to the forced queue, then after that, the forced queue cannot accept inputs that are blacklisted by the queued but not yet enforced policy. Without this extra check, Alloy tells us that we can reach a state when the head of the forced queue is blacklisted, which effectively means that finalized state becomes immutable. 

\begin{lstlisting}[language=alloy,style=mystyle]
no (L1.blacklist & c.diff.block_inputs.elems) 

all x : ForcedBlacklistPolicy, y : ForcedInput | 
    some L1.forced_queue 
 and x in L1.forced_queue.elems 
 and y.tx in L1.forced_queue.elems.tx 
 and y.tx in c.diff.block_inputs.elems  
      implies L1.forced_queue.idxOf[y] < L1.forced_queue.idxOf[x]
\end{lstlisting}

\point{Blacklisting properties}

\bp{Non-blacklisted Finalization} If an input is finalized, then it is not in the blacklist.
\begin{lstlisting}[language=alloy,style=mystyle]
always(
 all x : Input | 
  x in L1.finalized_state'.elems.block_inputs.elems 
 and x not in 
  L1.finalized_state.elems.block_inputs.elems
 implies x not in L1.blacklist
)
\end{lstlisting}
\bp{Forced Queue Integrity under Censorship} If a censored input is at the head of the forced queue, then the finalized state will never change.
\bp{Head Position Security} It never happens that head position of the queue is blacklisted.
\bp{Future Policy Compliance} All inputs that follow a new blacklist policy in the forced queue are not blacklisted by it.
\begin{lstlisting}[language=alloy,style=mystyle]
always (
  all x : ForcedBlacklistPolicy, y : ForcedInput |
     x in L1.forced_queue.elems
     and y in L1.forced_queue.elems
    and L1.forced_queue.idxOf[x] < L1.forced_queue.idxOf[y]
    implies y.tx not in x.predicate
)
\end{lstlisting}
\bp{Following Active Policy} If an input got forced and there is no queued blacklisting policy, then the input is not in the \texttt{L1.blacklist}.
\begin{lstlisting}[language=alloy,style=mystyle]
always (
  all y : ForcedInput |
     no L1.forced_queue.elems & ForcedBlacklistPolicy
     and y in L1.forced_queue.elems
    implies y.tx not in L1.blacklist
)
\end{lstlisting}
Given the properties outlined above, our model provides the following guarantee for users.

\highlightbox{Integrity of Processed Inputs under Blacklisting Mechanism}{
For any input processed in the L2 using a forced queue with blacklisting, the following conditions hold:
1) Any input that is processed is not blacklisted at the time of its processing.
2) When the input from the forced queue reaches the head of the queue, it is not blacklisted by the active policy.
}

\subsection{Forced Queue and Upgradeability}
\label{sec:upgradeability}
So far, we have discussed a basic model and presented mechanisms for forced queues and safely implementing a blacklist. In this section, we formalize a mechanism for applying upgrades to L2 contracts, noting that these upgrades could update the verifier key, indicate changes to the ZKP circuits, and even implement critical EIPs~\footnote{Ethereum Protocol Improvements (\url{https://github.com/ethereum/EIPs})} or other upgrades. Furthermore, these upgrades could alter the mechanisms previously presented. Hence, it is crucial that users have the opportunity to exit before an upgrade is enforced. To the best of our knowledge, we are the first to present a complete formalization and analysis of such a mechanism for rollups.

The upgrade mechanism operates in the following manner: 
\begin{enumerate}
    \item First, the upgrade is announced, including all relevant details, and this announcement is posted on L1. 
    \item Following the announcement, the system collects forced inputs for a specified duration.
    \item Once this duration elapses, meaning the timeout is reached, the rollup ceases to accept forced events and begins processing the forced queue until it is cleared.
    \item Once the forced queue is cleared, the upgrade is deployed, and the rollup shifts to a new operational mode. During this timeout period, only forced queue events are processed, which pressures the L2 operator to handle the forced queue swiftly. Halting the acceptance of forced transactions risks the assurance of a permissionless exit via the forced queue mechanism. Note that L2 users can still submit transactions through the L2 and receive pre-confirmations, but these cannot be finalized on the L1 until the forced queue is cleared.
\end{enumerate}
 In Alloy, we cannot reason about arbitrary system upgrades, so we instantiate the described upgradeability mechanism to update the blacklisting policy. The resulting mechanism is not equivalent to the mechanism described in Section~\ref{sec:blacklisting}.


\point{Comparison to blacklisting directly through the forced queue} 
In Section~\ref{sec:blacklisting}, we implemented a model which guarantees that inputs that were forced prior to the update of the blacklist policy will be finalized before the new blacklisting policy is deployed. However, every forced input which appears in the queue after the new blacklist policy must respect it. 

In this section, we describe a mechanism that updates the blacklisting policy through the generic upgradeability mechanism. In this case, users who observed the upgrade announcement can force their inputs (even if blacklisted by the pending upgrade) before the timeout. The time window between the upgrade announcement and the timeout could be especially useful when users disagree with the upgrade and would like to take action (e.g., quit the system) before the upgrade is deployed. At the same time, such a mechanism gives an opportunity to adjust to the bad actors.

In summary, both blacklisting strategies have their trade-offs. More specifically, the blacklisting strategy through the forced queue (see Section~\ref{sec:blacklisting}) 
allows the ZK-Rollup to immediately prevent censored transactions from being added into the forced queue (however, it cannot forbid execution of the transactions that are already present in the queue). On the other hand blacklisting strategy through the upgradeability mechanism allows users to notice the upgrade and quit the ZK-Rollup within the predefined window of time without receiving a damage.

\point{Data model extensions}
To support upgrades, we introduce some new types. The first is the new abstract type \texttt{UpgradeAnnouncement}, and its concrete subtype \texttt{BlacklistUpdateAnnouncement}, which includes a \texttt{blacklist\_policy}. This type represents an announcement of an upgrade, detailing the changes that will be applied, and users can inspect it as it will be submitted to L1. We also introduce another type, \texttt{Timeout}, associated with an \texttt{UpgradeAnnouncement}, indicating when the upgrade waiting period expires and it is time to enforce the upgrade.

\begin{lstlisting}[language=alloy,style=mystyle]
var abstract sig UpgradeAnnouncement {}

var sig BlacklistUpdateAnnouncement extends UpgradeAnnouncement {
  var blacklist_policy : one ForcedBlacklistPolicy
}

var sig Timeout {
  var upgrade : one UpgradeAnnouncement
}

one sig L1 {
    var ongoing_upgrade : lone UpgradeAnnouncement
    // + previously introduced fields
}
\end{lstlisting}

\point{Processing logic}
Handling upgrades requires three new predicates and amendments to \texttt{rollup\_process} and \texttt{receive\_forced}. First, \texttt{upgrade\_init} ensures there is no ongoing upgrade (line 2), sets the ongoing upgrade in the L1 contract (line 3), and verifies that this upgrade is not from the past (line 4).

\begin{lstlisting}[language=alloy,style=mystyle]
pred upgrade_init[f : UpgradeAnnouncement] {
  L1.ongoing_upgrade = none
  L1.ongoing_upgrade' = f
  (f not in Timeout.upgrade)
}
\end{lstlisting}

Next, the \texttt{upgrade\_timeout} predicate ensures that the current timeout corresponds to the ongoing upgrade. The \texttt{upgrade\_deploy} predicate then enforces the ongoing upgrade changes by applying the associated blacklist (if any) and setting the ongoing upgrade to None, which effectively updates the contracts and performs any modifications to the L2's logic.

Furthermore, we update two existing predicates. Specifically, for \texttt{rollup\_process}, we change the logic so if an upgrade is in process, there must be something in the forced queue to process (otherwise, L2 might delay the finalization of the upgrade indefinitely and keep operating with a stopped forced queue). This is enforced with the following line:

\begin{lstlisting}[language=alloy,style=mystyle]
upgrade_in_progress implies upgrade_forced_queue_processing
\end{lstlisting}
Similarly, in \texttt{receive\_forced}, we enforce that if an upgrade is in progress, new forced inputs are accepted only if the timeout for queuing has not yet occurred.

\point{Upgradeability Properties (UP)}
\up{Consistency of Upgrade Announcement, Timeout, and Enforcement} 
If an upgrade is deployed (policy changed), it must be preceded by a properly announced upgrade and a timeout period. The following Alloy code validates this sequence, confirming that any changes to the blacklist result exclusively from an upgrade process that has been announced and timed correctly. Specifically, the code stipulates that a change in the blacklist (line~2) must trace back to an Upgrade Announcement (line~4-8) that specifies the new blacklist settings. It also ensures that this change only takes effect following a \texttt{Timeout} associated with the announcement (lines~9--10), which verifies the waiting period was respected before the upgrade was enforced. 
\begin{lstlisting}[language=alloy,style=mystyle]
always (
  all is : set Input | L1.blacklist = is and 
  not L1.blacklist = L1.blacklist'  implies 
       some x : UpgradeAnnouncement | L1.ongoing_upgrade = x and
        once (L1.ongoing_upgrade = none and
              L1.ongoing_upgrade' = x  and
              L1.blacklist = is and
              (no { t : Timeout | t.upgrade = x })
        and ((some t : Timeout | t.upgrade = x)  
               releases L1.blacklist = is))
)
\end{lstlisting}

\up{Finalization of Forced Inputs before the Upgrade} 
After an upgrade, it is essential that all forced inputs are finalized to ensure that no transactions are left pending. The Alloy code below enforces this requirement by checking that if there is a change in the blacklist (indicative of an upgrade), then in the past, every forced input that was in the queue must eventually be finalized. This is achieved by ensuring that each transaction in the forced queue, at any point before the blacklist change, eventually becomes part of the finalized state. The code ensures that all pending transactions are processed before any announced upgrade takes effect.
\begin{lstlisting}[language=alloy,style=mystyle]
always (
 (not L1.blacklist = L1.blacklist') implies
  historically (
   all f : ForcedInput | 
    f in L1.forced_queue.elems implies
     eventually (f.tx in 
      L1.finalized_state.elems.block_inputs.elems
    )
  )
)  
\end{lstlisting}

\up{Post-Upgrade System Integrity} If the policy changes, then no ongoing upgrade is happening (the forced queue is unlocked, and the rollup process is unlocked).

\up{Consistency of Blacklisting During Upgrades} As long as the upgrade is ongoing, the \texttt{L1.blacklist} does not change.

Given all of the above properties, we arrive at the following:

\highlightbox{Upgrade Transparency and Guaranteed Action Before Upgrade Enforcement}{
An upgrade must begin with an announcement being published on L1 and users have a time window to submit their inputs before the upgrade is deployed; the mechanism also ensures that all inputs submitted to the forced queue during the time window are finalized before the upgrade is deployed.
}
This mechanism ensures that users are not caught off-guard by upgrades and are guaranteed to be able to act upon upgrade (e.g., exit from L2 rollup) before the upgrade is enforced.

\begin{figure*}[t]
    \centering
    \includegraphics[width=0.8\textwidth]{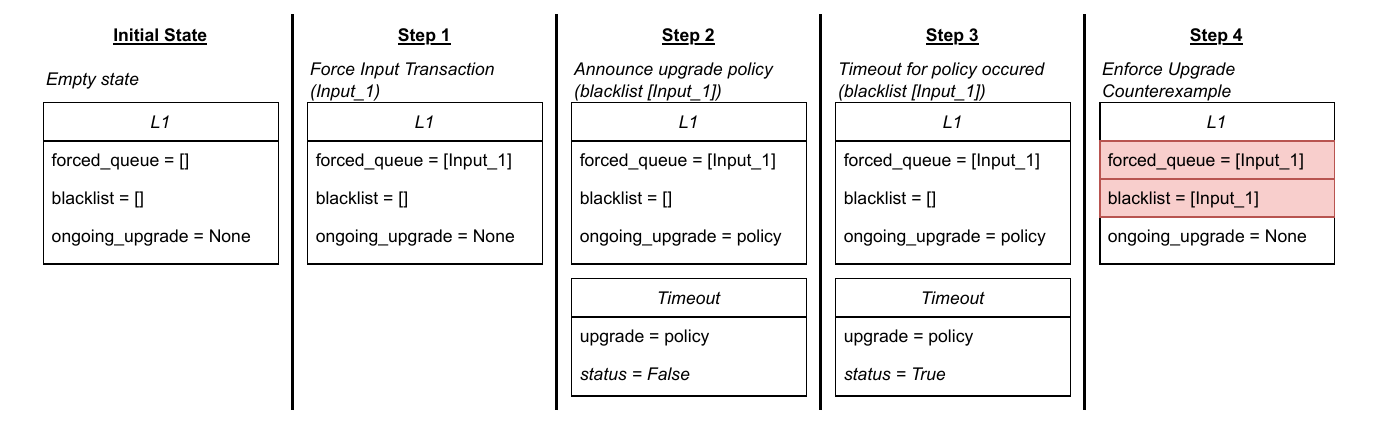}
    \caption{Counter-example generated by Alloy, illustrating potential issues with the naive upgradeability mechanism. In this scenario, the blacklisting policy targets \textsf{Input\_1}, leading to an L2 freeze because an input in the \textsf{forced\_queue} cannot be processed. This occurs because the mechanism does not require processing all transactions from the forced queue prior to executing an upgrade. 
}
    \label{fig:upgrade_counterexample}
\end{figure*}

\section{Leveraging Alloy Models for Testing Rollups}
\label{sec:leveragingalloy}

This section presents a complete methodology for testing and verifying the correctness of ZK-Rollup designs and implementations.
First, in Section~\ref{sec:testing}, we describe how our Alloy models can be used for design-time testing, enabling early detection of security and censorship resistance issues in the system design.
Then, in Section~\ref{sec:endtoend}, we explain how the formal properties we specify in Section~\ref{sec:alloy} (i.e., FQPs, BPs, and UPs) can be adapted for code-level verification and testing of smart contracts, using industry-standard formal verification and property-based testing frameworks.
Further, we demonstrate this methodology for Scroll's forced queue, where we adapt our Alloy model to reflect Scroll's architecture and translate the formal properties into Foundry invariants. This allows us to apply property-based testing directly on Scroll's smart contracts, bridging design-time specifications with implementation-level testing.
Together, these two layers of testing and verification form a practical end-to-end methodology for testing the correctness of ZK-Rollup security mechanisms from the initial design to the deployed code.

\point{Note on state-of-the-art rollups}
Currently, most general-purpose ZK-Rollups lack implementations of essential mechanisms like forced queues and delayed upgrades, which are crucial for inheriting L1's censorship resistance and security. 
The only exception is Scroll that implements a timeout-based forced queue that we thoroughly analyze in this section.
Yet, our approach can guide practitioners in testing their implementations as these mechanisms are developed. 

\subsection{Design-time Testing}
\label{sec:testing}
Building on the implementation details and security properties described in Section~\ref{sec:alloy}, we explore how our Alloy models can be used for design-time testing of ZK-Rollups. We provide an infeasibility result and highlight a vulnerability in the design of upgradeability mechanisms for ZK-Rollups. Note that this is not affecting any live systems as the safe upgraedability functionality has not be deployed by any ZK-Rollups.

\point{Scenario testing}
To validate our models, we implement several scenarios that test their functionality under various conditions, such as handling simultaneous commitments and proofs, processing in different sequences, and ensuring that commitments for an outdated state are not processed. These scenarios help demonstrate the model's capabilities and expose potential vulnerabilities. Also, it helps to detect incorrect specifications, e.g.:

\begin{lstlisting}[language=alloy,style=mystyle]  
run {
 eventually(
    some x : Input |  
        x in L1.blacklist 
    and (x not in all_finalized_inputs) 
  and eventually (x in all_finalized_inputs))
} for 7
\end{lstlisting}
In this scenario for the rollup with upgradeability and blacklisting (Section~\ref{sec:upgradeability}), we start with an empty blacklist state and no queued commitments or proofs. We assert that, eventually, we would like to reach a state where
there is an input \verb;x; which is blacklisted but not finalized. After that we would like to eventually reach a state when the input \verb;x; is finalized. This scenario is only possible if the rollup updates its blacklists two times where the first update blacklists \verb;x; and the second update removes \verb;x; from the blacklist and at the end \verb;x; gets finalized. When we execute the \href{https://alloy.readthedocs.io/en/latest/language/commands.html#run}{\texttt{run}} command, the Alloy Analyzer looks for models with up to seven instances of each signature that would satisfy the predicate. The Alloy Analyzer finds a satisfying instance that consists of a trace of length eight.

\point{Infeasibility of on-the-spot blacklisting}
When processing a blacklisting policy update transaction (Section~\ref{sec:blacklisting}), transactions that precede it in the queue must be processed. This means that otherwise censored transactions would have to be processed. If blacklisting is applied immediately, it  can generate a trace that gets the system stuck, breaking the guarantees of our system. To maintain the integrity of the forced queue, blacklisting should be enforced either through the forced queue (Section~\ref{sec:blacklisting}) or upgradeability mechanism (Section~\ref{sec:upgradeability}), ensuring correct operation of the forced queue. Moreover, the described upgradeability mechanism also ensures that all users have a fair chance to process their transactions before new policies take effect.
Notably, if we allow the query that updates the blacklisting policy ``on-the-spot'', then Alloy immediately generates simple counterexamples to most of the properties mentioned above.
Most importantly, due to the non-deterministic order of the actions in the system, ``on-the-spot'' updates can always lead to a situation where forced queue inputs get blacklisted, which makes the system stuck (i.e., no more blocks could be finalized). 

\point{Potential vulnerability in upgradeability policy}
The current state-of-the-art upgradeability design, often described as the safest approach, typically includes one week before enforcing an upgrade.\footnote{\url{https://medium.com/l2beat/introducing-stages-a-framework-to-evaluate-rollups-maturity-d290bb22befe}} Upgrades based solely on time can lead to situations where users who wish to exit the L2 before the upgrade are locked in because their transactions cannot be processed promptly. However, our solution (Section~\ref{sec:upgradeability}) mitigates this issue by ensuring that all transactions in the forced queue are processed before the upgrade takes effect, thus maintaining users' safety in case of malicious upgrades. Here, we discuss the potential pitfalls of relying solely on a timeout period for upgrades and then provide a counterexample demonstrating the issues with that mechanism, as produced by the Alloy Analyzer. 

Two issues could arise from merely enforcing the upgrade after a set period. For the first example, consider the case where we use a forced queue and upgradeability described in Section~\ref{sec:upgradeability}. However, we deploy the upgrade based solely on the timeout (i.e., not taking care of locking and emptying the forced queue before deploying the upgrade). Figure~\ref{fig:upgrade_counterexample} demonstrates the issue that might arise in that case produced by Alloy. Specifically, what happens is that first, we start with an empty forced queue, and then a user adds an input to the queue (\texttt{Input\_1}). Next, an announcement occurs, and then the timeout of that announcement is reached. This enforces the upgrade, updating the blacklist to include \texttt{Input\_1}. The result is that L2  becomes frozen because the head (\texttt{Input\_1}) of the queue is blacklisted.
Another problem could occur if the upgrade involves changing the functionality of the rollup and a user might submit a transaction to be forced before the upgrade because they disagree with the change and want to exit before the upgrade. However, because other transactions are in front of it (whether benignly or maliciously added), their transaction will not be processed. Additionally, the L2 might censor the user's transactions, preventing the user from exiting before the upgrade. 

These examples demonstrate severe issues with existing designs. Notably, such an issue existed in the zkSync Era's design (\emph{not in production}, as this mechanism has not been deployed yet), and we have worked closely with them to adapt their design to fix the issue. Our model presented in Section~\ref{sec:alloy} is immune to the described attacks. 
Finally, in Appendix~\ref{app:measurements}, we provide detailed statistics on our alloy model and a detailed analysis of the performance and scalability of our model.

\begin{figure}[t]
    \centering
    \begin{subfigure}[t]{0.48\textwidth}
        \centering
\begin{lstlisting}[basicstyle=\ttfamily\scriptsize]
rule guaranteedProcessing(env e) { 
    require forcedQueue.length > 0; 
    uint256 tx = forcedQueue[0]; 
    finalizeState(e); 
    assert !forcedQueue.contains(tx); 
}
\end{lstlisting}
        \caption{Certora rule encoding FQP1 for formal verification.}
        \label{fig:certora_fqp1}
    \end{subfigure}
    \hfill
    \begin{subfigure}[t]{0.48\textwidth}
        \centering
\begin{lstlisting}[basicstyle=\ttfamily\scriptsize]
function testGuaranteedProcessing() public { 
    vm.assume(forcedQueue.length > 0); 
    uint256 tx = forcedQueue[0]; 
    finalizeState(); 
    assert(!forcedQueue.contains(tx)); 
}
\end{lstlisting}
        \caption{Foundry (Forge) test encoding FQP1 as a property-based invariant.}
        \label{fig:foundry_fqp1}
    \end{subfigure}
    \caption{Example of using Certora (a) and Foundry (b) to express the \textbf{Guaranteed Processing} property (FQP1).}
    \label{fig:certora_foundry_fqp1}
\end{figure}

\subsection{Code-level Verification/Testing}
\label{sec:endtoend}

In the previous section, we demonstrated how our formal model aids in the high-level design verification of ZK-Rollup mechanisms. Here, we propose an extension of this approach to the implementation level (smart contracts), illustrating how the properties defined in Section~\ref{sec:alloy}, i.e., Forced Queue Properties (FQPs), Blacklisting Properties (BPs), and Upgradeability Properties (UPs), can be employed to test actual smart contract implementations of these mechanisms.

\point{Formal verification and property-based testing of smart contracts}
Formal verification tools like Certora~\cite{certora2025whitepaper} and property-based testing frameworks such as Foundry~\cite{foundry2021} and Echidna~\cite{grieco2020echidna} are widely adopted in the DeFi ecosystem, with protocols like Lido and AAVE utilizing them to ensure contract correctness~\cite{chaliasos2024smart}. Certora allows developers to write specifications in the Certora Verification Language (CVL), enabling the formal verification of smart contracts against defined properties. Foundry and Echidna facilitate property-based testing by generating a wide range of inputs to test contract behaviors against specified invariants.

\point{End-to-end analysis of ZK-Rollups} Our methodology involves adapting the formally verified properties from our Alloy model (FQPs, BPs, and UPs) to the smart contracts of a selected rollup. These adapted properties are then verified/tested using frameworks like Certora or Foundry. Given that Alloy has formally verified these properties, any counterexamples found by the employed tool indicate implementation bugs. This approach ensures the implementation aligns with the formally verified design, bridging the gap between high-level specifications and practical deployment.

To illustrate, consider \textit{FQP1} (Section~\ref{sec:forcedqueue}). This property ensures that if the forced queue is non-empty and a new state is finalized, then the head of the forced queue must be processed and removed. For simplicity, consider a smart contract of a rollup with a queue \texttt{forcedQueue} containing transaction IDs. Figure~\ref{fig:certora_fqp1} depicts the rule for Certora, while Figure~\ref{fig:foundry_fqp1} shows the invariant for Foundry. This example demonstrates how Alloy properties can be translated into formal verification rules or invariants, ensuring that implementations adhere to their specifications. In Appendix~\ref{app:extensions}, we explore more potential future work of automating the whole process. 

\point{Analyze Scroll's forced queue}
Scroll implements only the forced queue mechanism from the mechanisms presented in this paper. Unlike our general model, Scroll's implementation enforces the forced queue only when a timeout occurs, i.e., when the head of the queue has been delayed beyond a predefined threshold. At that point, the rollup cannot progress until it processes the timed-out forced queue messages. To accurately capture this behavior, we extended our Alloy model by refining the forced queue properties (FQP) to encode timeouts and introducing four Scroll-specific properties: (i) Rolling hash integrity to ensure correct hashing of queued messages, (ii) Enforced mode activation to verify that enforced mode is triggered after a timeout, (iii) Mode consistency to guarantee mutual exclusivity between enforced and normal modes, and (iv) Fee payment to confirm that forced queue messages cover their required fees.

We translated this adapted Alloy model into a property-based testing suite using Foundry, implementing all specified properties in approximately 1,000 lines of Solidity code. Each property was encoded as an invariant checked after random operations, with a total of 10,000 runs. No violations were detected in the forced queue implementation of Scroll. The full adaptation and testing process took three days, aided significantly by Scroll's existing support for Foundry-based unit tests and mock operations such as batch proving and verification. The existing infrastructure was a key reason for selecting Foundry over Certora.

\section{Discussion}
\label{sec:discussion}

\point{Alloy's Bounded Scope and Security Guarantees}
Alloy exhaustively explores all behaviours \emph{within} a bound on object count and trace length; success inside that window does \emph{not} prove universal correctness~\cite{brunel2021alloy6}. It is therefore a powerful bug-finder, but not a complete verifier. We ran the analyser with at most 10 objects (scope) and 10-step traces. This setting follows the \emph{small-scope hypothesis}, which holds that most design errors admit small counter-examples~\cite{jackson1996elements,jackson2019alloy}. The flaw in Fig.~\ref{fig:upgrade_counterexample} arises with a single forced transaction and one upgrade event; adding a second queued transaction reveals the subtler `blocked-exit' scenario.
To gauge this approach, we injected five synthetic faults: four surfaced at scope~$\le 5$, the fifth at scope~8. This suggests a scope of 8 would suffice, yet we conservatively used up to 10. Appendix~\ref{app:measurements} shows how SAT-solver time grows exponentially with scope, reinforcing the need to cap it. Importantly, our model may miss issues that necessitate a larger scope to be manifested.

\point{Analysing Further L2 Mechanisms}
Our model focuses on mechanisms enforced by \emph{L1} contracts: forced queues, blacklisting, and upgradeability, because these must remain correct regardless of how the L2 is engineered; they are essential for a rollup to inherit the security properties of its base chain. Other mechanisms that touch multiple layers have not been analyzed. Withdrawals, for instance, interact with both layers: an exit begins on L2 but is finalised on L1 only after the corresponding L2 transaction is settled, so modelling them faithfully would require incorporating the bridge contract and relevant L2 logic, an extension we leave to future work that can build on prior bridge analyses~\cite{lee2023sok}. Further, data-availability guarantees, which depend on blob-space economics or external DA layers, introduce complexities beyond our present scope. In short, we prioritise L1-relevant mechanisms that must remain sound even under a hostile L2; extending the model to cross-layer withdrawals is a natural next step, while DA and sequencer-level protocols will likely require different analyses.

\point{Deployment Status and Practical Relevance of ZK-Rollups}
ZK-Rollups are no longer research prototypes: according to \texttt{L2Beat} (July 2025), six of the ten largest rollups by TVL are ZK-Rollups, together securing well over \$3 billion in user assets\footnote{\url{https://l2beat.com/scaling/summary}}. The remaining four are optimistic rollups, yet even optimistic rollups could potentially move towards ZK proofs. For example, OP-Succinct (Succinct Labs) aims to let existing optimistic stacks switch to validity proofs with minimal code changes.~\footnote{\url{https://github.com/succinctlabs/op-succinct}} Of the three mechanisms we formalize, \emph{forced exits} and \emph{timelocked upgrades} are already on every major rollup's public roadmap; nevertheless, only Scroll has a deployed forced-queue today (c.f., Section~\ref{sec:endtoend}). The \emph{safe-blacklisting} mechanism is less ubiquitous but increasingly demanded by institutionally focused chains that require compliance; our analysis shows that simple `on-the-spot' blacklists could break liveness (Section~\ref{sec:testing}), while our model preserves user guarantees. In conclusion, ZK-Rollups are widely deployed, the proposed mechanisms are on the critical path for all of them, and the presented formal model has already proven its practical value by uncovering potential issues in the interactions between mechanisms and helping test critical properties of a live system.

\section{Related Work}
\label{sec:related}

\point{Formal Methods for Web Security}
Our work is inspired by the seminal work of Akhawe et al.~\cite{akhawe2010towards} on applying formal modeling to analyze the security of web mechanisms and applications. Specifically, this work provides a formal model of the web platform to dissect the security of various web mechanism designs using Alloy. Their analysis included the specification of threat models tailored for evaluating the security of web platforms. Analogously, we crafted our analysis around the strongest conceivable adversary (i.e., a malicious L2), examining rollup designs under this threat model. Moreover, Akhawe et al.~\cite{akhawe2010towards} established a set of broadly applicable security goals and analyzed various security properties for web applications. Similarly, we defined fundamental security goals tailored for rollup users, defined security properties, and systematically evaluated different scenarios using Alloy. For those interested in formal methods for security, the recent survey by Kulik et al.~\cite{kulik2022survey} offers a detailed examination of formal methods accessible to designers of security-critical systems. 

\point{Formal methods in blockchain and ZKPs}
There is a growing body of work verifying different aspects of blockchain, from smart contracts to consensus algorithms and ZK proof generation. We mention only some relevant work here.
For smart contract verification, existing research endeavors primarily aim to demonstrate the absence of common vulnerabilities or rely on user-defined specifications to verify correctness. Notable surveys by Murray and Anisi~\cite{murray2019survey} and Garfatta et al.~\cite{garfatta2021survey} offer comprehensive overviews of formal verification methods tailored for smart contracts. However, it is important to note that our work diverges from this trajectory, focusing instead on providing assurances regarding the design of mechanisms implemented in smart contracts that operate rollups.

On the front of formal verification for Byzantine fault tolerance (BFT) in blockchain systems, seminal works by Tholoniat and Gramoli~\cite{tholoniat2022formal} and Yoo et al.~\cite{yoo2019formal} have contributed significantly. 
Regarding ZKPs, recent efforts have explored the verification of verifiers for ZK-rollups~\cite{ouroboros}. Additionally, several initiatives have focused on formally proving the absence of common bugs in ZK circuits~\cite{chaliasos2024sok}. In contrast to those efforts, we focus on the design of mechanisms that are complementary to ZKPs.

\point{Rollup Security}
Prior works have focused on qualitatively analyzing optimistic and ZK-Rollups. Gorzny et al.~\cite{gorzny2022ideal} established a wishlist of properties that an escape hatch mechanism (also known as a forced queue) should possess to be considered trustworthy. Their paper focuses on qualitative characteristics that escape hatches should exhibit, such as executing arbitrary transactions, being supported by default in a rollup, and efficiency. In contrast, our paper formally defines the security properties of escape hatches and analyzes various scenarios using Alloy, enabling the identification of vulnerabilities in flawed designs. 

In their recent work, Koegl et al.~\cite{koegl2023attacks} presented nine main attack threats for rollups, briefly describing potential attacks in both optimistic and ZK-Rollups, such as censorship attacks, DoS attacks, and client vulnerability risks. Our work extends some of these potential attacks by formally modeling ZK-Rollups, presenting a clear and strong adversary, and outlining the security properties necessary to ensure a secure ZK Rollup design. 
Finally, complementing our work, Gorzny and Derka~\cite{gorzny2024rollup} presented a framework to qualitatively evaluate existing rollups based on different dimensions: familiarity, finality time, modularity, and maturity. 

In comparison to the aforementioned works, we are the first to formally specify the main properties of ZK-Rollups and analyze various designs using Alloy, providing formal guarantees of the security of specific ZK rollup designs. In essence, while prior works provide generic qualitative directions, we have developed the first tooling based on formal methods to help practitioners identify potential issues in the security-critical mechanisms of rollups.

\section{Conclusion}
\label{sec:conclusions}
Rollups are heralded as the foremost solution to blockchain scalability, collectively securing over~\$30 billion in value. However, recent incidents have led to censorship and halting block production on rollups. Despite these astronomical monetary amounts, the security properties of rollups remain inadequately understood.

Our work introduces an extensible formal model for ZK-Rollups that facilitates the design and evaluation of rollups' critical mechanisms and enhances the understanding of their security properties. Using Alloy, a language used for modeling complex systems for over a decade, we create models accessible to developers, security auditors, and researchers, enabling automatic reasoning with rapid results. 
We further show how the security properties defined in Alloy can be used to verify or test the smart contracts implementing rollup logic on-chain, using formal verification or property-based testing. Our primary objective is to lay the foundation for creating and advancing secure L2 solutions that are verifiably proven.

\begin{acks}
We would like to thank the anonymous CCS reviewers for their insightful feedback on previous versions of the paper. D. Firsov was supported by the Estonian Research Council grant no. PSG749.
\end{acks}

\bibliographystyle{ACM-Reference-Format}
\bibliography{sample-base}

\appendix

\section{Alloy Measurements}
\label{app:measurements}

\begin{figure}[h]
\centering
\resizebox{.35\textwidth}{!}{%
\begin{tabular}{@{}lrrr@{}}
\toprule
\multirow{2}{*}{\textbf{Mechanism}} & \textbf{Lines of} & \textbf{No. of} & \textbf{Solve time} \\
                   & \textbf{code} & \textbf{Clauses}    &  \textbf{(sec)}    \\
\midrule
Simple             & 295 & 1,814,483 & 63.603  \\
Forced             & 529 & 3,633,032 & 89.623  \\
Blacklist          & 721 & 5,461,695 & 152.688 \\
Upgrade            & 970 & 6,149,438 & 125.703 \\
Upgrade+Blacklist  & -   & 7,978,101 & 188.768 \\
\bottomrule
\\
\end{tabular}
}
\caption{Various cumulative statistics for each rollup mechanism. The No. of Clauses and Solve time are for scope 5 and 10 steps.}
\label{tab:performance_metrics}
\end{figure}

\begin{figure}[h]
    \centering
    \includegraphics[width=0.35\textwidth]{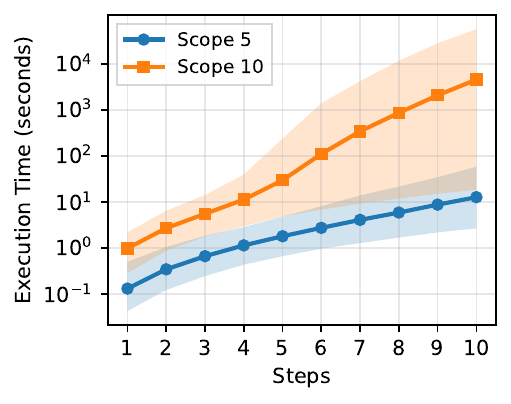}
    \caption{Log-scale graph of analysis time for 5 and 10 scopes. The colored part depicts the variance across properties.}
    \label{fig:time_per_scope}
\end{figure}

In this paper, we use Alloy Analyzer~6 to verify the invariants and assertions of various L2 mechanisms. This involved adapting our model for each ZK-Rollup mechanism and then systematically testing them within the Alloy environment. The scope for our tests was set up to allow us to specify more fine-grained bounds on each type, but it does not imply any limitation on the overall number of system states that can be explored. All experiments were performed on a MacBook Pro Apple M1 Max CPU with ~32 GB of RAM.

For each case, we quantified the number of new lines of Alloy code written, the total number of clauses generated by the analyzer, and the time to solve these clauses using an SAT solver. Figure~\ref{tab:performance_metrics} shows the lines of code required for each mechanism, the cumulative number of clauses of the properties of each mechanism, and the cumulative solve time in seconds for scope 5 and steps 1--10. The complexity and the time required to solve the clauses increase with more complex rollup mechanisms.
Figure~\ref{fig:time_per_scope} depicts the running time for different step values for scopes 5 and 10, demonstrating a significant increase in the SAT solver time as the scope increases, as well as the steps increase. The observed exponential growth is typical of SAT-based analyses. Finally, the colored area demonstrates the variance between checking different properties.

\section{Model Extensions}
\label{app:extensions}

In this section, we propose some potential extensions that can be built on top of our model and could further enhance the state of ZK-Rollups, by making our model more practical, modeling more functionalities, and capturing different implementation details.

\point{Enforcing Processing from the Forced Queue in Intervals}
While our model requires that every new block processed must include a transaction from the forced queue if it is not empty, this might be overly strict for practical implementations due to the asynchronous nature of blockchain systems. A more practical approach would be to enforce the processing of the head of the forced queue every~3--5 blocks being finalized. This adjustment still offers the same guarantees to users regarding eventual processing, but allows for greater flexibility and efficiency in the system's operation.

\point{Modeling Optimistic Rollups}
Optimistic rollups, which rely on fraud proofs handled by the L1 for dispute resolution, present a more complex scenario than our current model. Extending our model to include optimistic rollups would involve formalizing the fraud proof mechanism. 

\point{Analyzing Withdrawal Windows in Optimistic Rollups}
Optimistic rollups that have been deployed generally employ a seven-day withdrawal period. Although numerous reasons exist for the necessity of this duration~\cite{rollup-window}, newer methods propose shortening these periods. Nevertheless, the potential security compromises of these more assertive suggestions~\cite{reducing-window} still require thorough examination.

\point{Consideration of State Diffs vs. Transaction Inputs}
To capture more granular changes within blocks, our model could be extended to differentiate between state diffs and complete transaction inputs. This would allow for a more detailed understanding of the state transitions within the rollups, aligning more closely with how changes are batched and processed in practical implementations. Nevertheless, this should not change the properties described in our model.

\point{Introducing Data Availability and Expiration}
Incorporating data availability and expiration into our model could significantly enhance its utility. This involves modeling how data is made available and the potential expiration or deletion of data, which is critical for maintaining the integrity and accessibility of the rollup data over time.

\point{Associating Finalized States with Hashes Instead of Actual Blocks}
To address the potential disappearance of data (data availability issues), our model could be extended to associate finalized states with hashes rather than actual block contents. This approach would ensure that the system can still verify the integrity of the data even if the actual data is no longer directly accessible.

\point{Detailed Modeling of L1 Blocks and Finality}
The model could be extended to include detailed modeling of L1 blocks and their finality,\footnote{\url{https://ethereum.org/en/roadmap/single-slot-finality/}} capturing how Ethereum's finality mechanisms, impact L2 operations. Formalizing how L1 reorgs affect L2s would provide critical insights into how to design L2 systems to handle these events safely.

\point{Modeling L2 Operations Beyond Smart Contracts}
Expanding our model to include specifications of L2 operations and components beyond just the smart contracts on L1 could provide a foundation for designing L2 sequencers and other critical infrastructure, such as the interactions between L2 sequencers and provers. This would build upon the provided L1 model, offering a comprehensive framework for the architectural design of L2 rollups.

In this work, we provide the fundamental missing formalization and implement it in Alloy to provide guarantees about its correctness. We believe that this base could be used as a framework to analyze the design of critical mechanisms of L2 rollups.

\end{document}